\documentclass[pre,showpacs,floats,twocolumn,floatfix,superscriptaddress]{revtex4-1}

\usepackage{mathrsfs}
\usepackage{bm,amsbsy,amssymb,amsmath}
\usepackage{caption}
\usepackage{subcaption}
\usepackage{graphics,graphicx,dcolumn,fleqn,epic,eepic,float,tabularx}
\usepackage{multirow,rotate,rotating,color}
\usepackage[utf8]{inputenc}
\newcommand{\figref}[1]{Fig.~\ref{fig:#1}}
\newcommand{\eqnref}[1]{Eq.~(\ref{eq:#1})} 
  \definecolor{tuered}{RGB}{214,0,74}
  \definecolor{tueblue}{RGB}{0,102,204}
  
  \newcommand{\revisedtext}[1]{#1}

\usepackage{tikz}
\usepackage{relsize}
\tikzset{fontscale/.style = {font=\relsize{#1}}}
\usetikzlibrary{calc}
\begin{document}
\title{Tunable Dipolar Capillary Deformations for Magnetic Janus Particles \\
 at Fluid-Fluid Interfaces}
 \author{Qingguang Xie}
  \email{q.xie1@tue.nl}
  \affiliation{Department of Applied Physics, Eindhoven University of Technology, P.O. Box 513, NL-5600MB Eindhoven, The Netherlands}
  \author{Gary B. Davies}
  \email{g.davies.11@ucl.ac.uk}
   \affiliation{Centre for Computational Science, University College London, 20 Gordon Street, London WC1H 0AJ, United Kingdom}
  \author{Florian G\"unther}
   \email{f.s.guenther@tue.nl}
    \affiliation{Department of Applied Physics, Eindhoven University of Technology, P.O. Box 513, NL-5600MB Eindhoven, The Netherlands}
  \author{Jens Harting}
  \email{j.harting@tue.nl}
  \affiliation{Department of Applied Physics, Eindhoven University of Technology, P.O. Box 513, NL-5600MB Eindhoven, The Netherlands}
  \affiliation{Faculty of Science and Technology, Mesa+ Institute, University of Twente, 7500 AE Enschde, The Netherlands}
\date{\today}

\begin{abstract}
Janus particles have attracted significant interest as building blocks
for complex materials in recent years. Furthermore, capillary
interactions have been identified as a promising tool for directed
self-assembly of particles at fluid-fluid interfaces. In this paper, we
develop theoretical models describing the behaviour of magnetic
Janus particles adsorbed at fluid-fluid interfaces interacting with an
external magnetic field. Using numerical simulations, we test the models
predictions and show that the magnetic Janus particles deform the
interface in a dipolar manner. We suggest how to utilise the resulting
dipolar capillary interactions to assemble particles at a
fluid-fluid interface, and further demonstrate that the strength of
these interactions can be tuned by altering the external field strength,
opening up the possibility to create novel, reconfigurable materials.
\end{abstract}

\maketitle

\section{Introduction}

Colloidal Janus particles have drawn special attention during the past two decades for their potential in materials science~\cite{Gennes1992}.
Janus particles are characterized by anisotropic surface chemical (e.g. wetting or catalytic) or physical (e.g. optical, electric, or magnetic) properties at well-defined areas on the particle.
This combination of chemical anisotropy and response to external fields makes Janus particles 
promising building blocks of reconfigurable and programmable self-assembled structures~\cite{Andreas2013,Yan2012,Smoukov2009, Ren2012, Ruditskiy2013}.

Janus particles strongly adsorb at fluid-fluid interfaces~\cite{Binks2001}, making the formation of 2-D structures accessible.
For symmetric Janus particles composed of hydrophobic and hydrophilic hemispheres, the equilibrium
contact angle is $90^o$ since each hemisphere immerses in its favourable fluid, and the interface remains flat~\cite{Ondurcuhu1990}.
However, due to surface roughness~\cite{Adams2008}, anisotropic shape~\cite{Park2012a,Park2012c}, or the influence of external forces, Janus particles can tilt with respect to the interface.
In a tilted orientation, the fluid-fluid interface around the Janus particle deforms in a dipolar fashion in order to fulfil boundary conditions stipulated by Young's equation~\cite{Rezvantalab2013}. 
Assuming small interface deformations, the particle-induced interface deformations obey $\nabla^2 h = 0$, where $h$ is the interface height,  
which can be solved using a multipolar analysis, analagous to 2D electrostatics~\cite{Stamou2000}.
These particle induced deformations, called capillary interactions, can cause particles to attract and repel in specific orientations, making them a useful tool for controlling the behaviour of particles at interfaces.

Previous investigations into capillary interactions between particles at fluid-fluid interfaces have focussed mainly around two themes: particle weight-induced deformations, 
which lead to monopolar interactions between particles and are responsible for e.g. the Cheerios effect~\cite{vella_cheerios_2005}; and surface roughness or shape anisotropy 
induced deformations, which lead to quadrupolar interactions between particles~\cite{Loudet2005} and are responsible for e.g. the suppression of the coffee ring effect~\cite{yunker_suppression_2011}.
With respect to Janus particles, Brugarolas et al.~\cite{Brugarolas2011a} showed that quadrupolar capillary interactions induced by surface roughness can be used to form fractal-like structures of Janus nanoparticle-shelled bubbles.
However, a significant limitation of the above mentioned capillary interactions is that they are not dynamically tunable because they depend on the particle properties alone.

Davies et al.~\cite{Gary2014a} recently found a way of creating dynamically tunable dipolar capillary interactions between magnetic ellipsoidal particles adsorbed at an interface under the influence of an external magnetic field. The structures that form depend on the dipole-field coupling, which can be controlled dynamically~\cite{Gary2014b}. However, it is also desirable to create tunable capillary interactions between spherical particles without relying on particle shape anisotropy. 

In this paper, we show how to create tunable dipolar capillary interactions
using spherical Janus particles adsorbed at fluid-fluid interfaces. The Janus
particles have a dipole moment orthogonal to their Janus boundary and are
influenced by an external magnetic field directed parallel to the interface.
The field causes the particles to experience a magnetic torque, but surface 
tension opposes this torque, and the particles therefore tilt with respect to the 
interface. When tilted, the particles deform the interface in a dipolar fashion. 

We develop a free energy based model of the behaviour of a single particle at the interface, and a model which takes into account small interface deformations. We numerically investigate the predictions of the models using lattice Boltzmann simulations and highlight that the dipolar interface deformations lead to novel and interesting particle behaviour at the interface. Finally, we explain how to use these interface deformations to manufacture tunable capillary interactions between many particles at a fluid-fluid interface, which could be utilised to assemble reconfigurable materials. 

This paper is organised as follows. We present our hybrid molecular dynamics-lattice Boltzmann simulation method in Section~\ref{sec:method}. In Section~\ref{sec:theory}, we develop two theoretical models describing the behaviour of Janus particles at fluid-fluid interfaces. Section~\ref{sec:simu} contains our simulation results, and we compare these results with our theoretical models from Section~\ref{sec:theory}. Finally, Section~\ref{sec:final} concludes the article.
\section{Simulation method}
\label{sec:method}
\subsection{The multicomponent lattice Boltzmann method}
\label{ssec:lb}
We use the lattice Boltzmann method (LBM) to simulate the motion of each fluid. 
The LBM is a local mesoscopic algorithm, allowing for efficient parallel implementations, and has demonstrated itself as a powerful tool for numerical simulations of fluid flows~\cite{Succi2001}.
It has been extended to allow the simulation of, for example, multiphase/multicomponent 
fluids~\cite{Shan1993,Cappelli2015} and suspensions of particles of arbitrary shape and wettability~\cite{ladd-verberg2001, Jansen2011, Gunther2013a}.

We implement the pseudopotential multicomponent LBM method of Shan and Chen~\cite{Shan1993} with a D3Q19 lattice~\cite{Qian1992} and review some relevant details in the following. For a detailed description of the method, 
we refer the reader to the relevant literature.~\cite{Jansen2011, Gunther2013a, Frijters2012, Gunther2013, Frijters2014}
Two fluid components are modelled by following the evolution of each distribution function discretized in space and time according to the lattice Boltzmann equation:
\begin{eqnarray}
  \label{eq:LBG}
  f_i^c(\vec{x} + \vec{c}_i \Delta t , t + \Delta t)=f_i^c(\vec{x},t) - \frac{\Delta t} {\tau^c} [  f_i^c(\vec{x},t)  \nonumber \\ 
   - f_i^\mathrm{eq}(\rho^c(\vec{x},t),\vec{u}^c(\vec{x},t))]
  \mbox{,}
\end{eqnarray}
where $i=1,...,19$, $f_i^c(\vec{x},t)$ are the single-particle distribution functions for fluid component $c=1$ or $2$, 
$\vec{c}_i$ is the discrete velocity in $i$th direction, and $\tau^c$ is the relaxation time for component $c$. 
The macroscopic densities and velocities are defined as  
$ \rho^c(\vec{x},t) = \rho_0 \sum_if^c_i(\vec{x},t)$, where $\rho_0$ is a reference density, and $\vec{u}^c(\vec{x},t) = \sum_i  f^c_i(\vec{x},t) \vec{c}_i/\rho^c(\vec{x},t)$, respectively.
Here, $f_i^\mathrm{eq}(\rho^c(\vec{x},t),\vec{u}^c(\vec{x},t))$ is a third-order equilibrium distribution function.
When sufficient lattice symmetry is guaranteed, the Navier-Stokes equations can be recovered from \eqnref{LBG} on appropriate length and time scales~\cite{Succi2001}.
For convenience we choose the lattice constant $\Delta x$, the timestep $ \Delta t$, the unit mass $\rho_0 $ and the relaxation time $\tau^c$ to be unity, which leads 
to a kinematic viscosity $\nu^c$ $=$ $\frac{1}{6}$ in lattice units.

The Shan-Chen multicomponent model introduces a mean-field interaction force 
\begin{equation}
  \label{eq:sc}
  \vec{F}_{\mathrm{C}}^c(\vec{x},t) = -\Psi^c(\vec{x},t) \sum_{c'}g_{cc'} \sum_{\vec{x}'} \Psi^{c'}(\vec{x}',t) (\vec{x}'-\vec{x})
\end{equation}
 between fluid components $c$ and $c'$~\cite{Shan1993}, 
in which $\vec{x}'$ denote the nearest neighbours of lattice site $\vec{x}$ and $g_{cc'}$ is a coupling constant determining the surface tension.
$\Psi^c(\mathbf{x},t)$ is an ``effective mass'', chosen with the following functional form:
\begin{equation}
  \label{eq:psifunc}
  \Psi^c(\vec{x},t) \equiv \Psi(\rho^c(\vec{x},t) ) = 1 - e^{-\rho^c(\vec{x},t)}
   \mbox{.}
\end{equation}
This force is then applied to the component $c$ by adding a shift $\Delta \vec{u}^c(\vec{x},t) = \frac{\tau^c \vec{F}_{\mathrm{C}}^c(\vec{x},t)}{\rho^c(\vec{x},t)}$ to the velocity $\vec{u}^c(\vec{x},t)$ in the equilibrium distribution.
The Shan-Chen LB method is a diffuse interface method, resulting in an interface width of $\approx 5\Delta x$~\cite{Frijters2012}.
\begin{figure*}
\begin{subfigure}{.4\textwidth}
\begin{tikzpicture}
	\fill[white,opacity=0.4] (-4,-3) rectangle (4.0,0.0);
	\fill[white,opacity=0.4] (-4,0) rectangle (4.0,3.0);
	
	\draw (0,-0.0) circle (1.75);
	\shade[ball color=yellow] (0,-0.0) circle (1.75);
	
	\draw[dashed] (-4,0) to (4,0);
		
	\node at (0,3.5) [fontscale=3] {$\mathbf{H}=0$};
	
	\draw[thick,red] ([shift=(0:0.5cm)]0.6,0) arc (0:55:0.5cm);
  	\node[white] at (1.4,0.25) [fontscale=2] {$\varphi$};

	\node at (-3.0,0.6) [fontscale=3] {Fluid 1};
	\node at (-3.0,-0.6) [fontscale=3] {Fluid 2};

 \filldraw[fill=red!60!white, draw=green!50!black]
    (0,0) -- (1.75,0.0) arc (0:180:1.75) -- (0,0);

 \filldraw[fill=blue!60!white, draw=green!50!black]
    (0,0) -- (-1.75,0.0) arc (180:360:1.75) -- (0,0);
    
    \draw[ultra thick, black,arrows=->] (0,0) to (0,2.5); 
    \draw[ultra thick, black] (-1.75, 0) to (1.75, 0); 
    \node at (-0.35,2.3) [fontscale=2] {$\mathbf{m}$};
    	\node at (1.3,1.0) [fontscale=2] {$Apolar$};
	\node at (1.3,-1.0) [fontscale=2] {$Polar$};
	\draw[dashed] (0,-1.75) to (0,2);
\end{tikzpicture}
\caption{Equilibrium orientation}
\label{fig:equi}
\end{subfigure}
\begin{subfigure}{.4\textwidth}
\begin{tikzpicture}

	\fill[white,opacity=0.4] (-4,-3) rectangle (4.0,0.0);
	\fill[white,opacity=0.4] (-4,0) rectangle (4.0,3.0);
	
	\draw[ultra thick,red] (1.5,0.75) to [out=330,in=180] (4,0.0);
	\draw[ultra thick,red] (-1.5,-0.75)  to  [out=150,in=360](-4,0.0);
	
	\draw (0,-0.0) circle (1.75);
	\shade[ball color=yellow] (0,-0.0) circle (1.75);
	
	\draw[dashed] (-4,0) to (4,0);
	
	\draw[ultra thick, black,arrows=->] (0.5,3.0) to (-0.5,3.0);
	\node at (0,3.5) [fontscale=3] {$\mathbf{H}$};
	
	\draw[ultra thick] (-1.60,-0.75) -- (1.60,0.75);
	\draw[ultra thick] (-1.0,-1.45) -- (1.0,1.45);

 \filldraw[fill=blue!30!white, draw=green!50!black]
    (0,0) -- (1.6,0.75) arc (25:55:1.75) -- (0,0);
 \filldraw[fill=red!60!white, draw=green!50!black]
    (0,0) -- (1.0,1.45) arc (55:205:1.75) -- (0,0);
    
 \filldraw[fill=red!30!white, draw=green!50!black]
    (0,0) -- (-1.6,-0.75) arc (205:235:1.75) -- (0,0);
    
   \filldraw[fill=blue!60!white, draw=green!50!black]
    (0,0) -- (-1.0,-1.45) arc (235:385:1.75) -- (0,0);
    
    	\draw[thick,black] ([shift=(0:0.5cm)]-0.5,0.6) arc (90:155:0.5cm);
  	\node[black] at (-0.5,0.8) [fontscale=2] {$\varphi$};
  	
  	\draw[ultra thick, black,arrows=->] (0,0) to (-1.87,1.29);
  	\draw[ultra thick, black] (-1.0,-1.45) -- (1.0,1.45); 
  	\draw[thick, black, dashed] (1.6,0.0) -- (1.6,0.75);  
  	\draw[dashed] (-4,0) to (4,0);
  	\node at (1.96,0.25) [fontscale=2] {$\zeta_1$};
  	\draw[dashed] (0,-1.75) to (0,1.75);
    	\node at (1.1,1.0) [fontscale=2] {$A_{p1}$};
	\node at (-1.1,-0.85) [fontscale=2] {$A_{a2}$};
	
	\node at (-2.1,1.5) [fontscale=2] {$\mathbf{m}$};

\end{tikzpicture}
\caption{Tilted orientation}
\label{fig:def}
\end{subfigure}
\caption{A single Janus particle adsorbed at a fluid-fluid interface in its
equilibrium orientation $(a)$ and in a tilted orientation $(b)$. The Janus
particle consists of an apolar and a polar hemisphere.  The particle's magnetic
dipole moment $\mathbf{m}$ is orthogonal to the Janus boundary, and the
external magnetic field, $\mathbf{H}$, is directed parallel to the interface.
The tilt angle $\varphi$ is defined as the angle between the particle's dipole moment and the undeformed interface normal.
$A_{a2}$ and $A_{p1}$ are the surface areas of the apolar hemisphere immersed in fluid $2$ and the polar hemisphere immersed in fluid $1$, respectively.
The bold red line represents the deformed interface and $\zeta_1$ is the maximal interface height at the contact line.
}
\label{fig:geo}
\end{figure*}
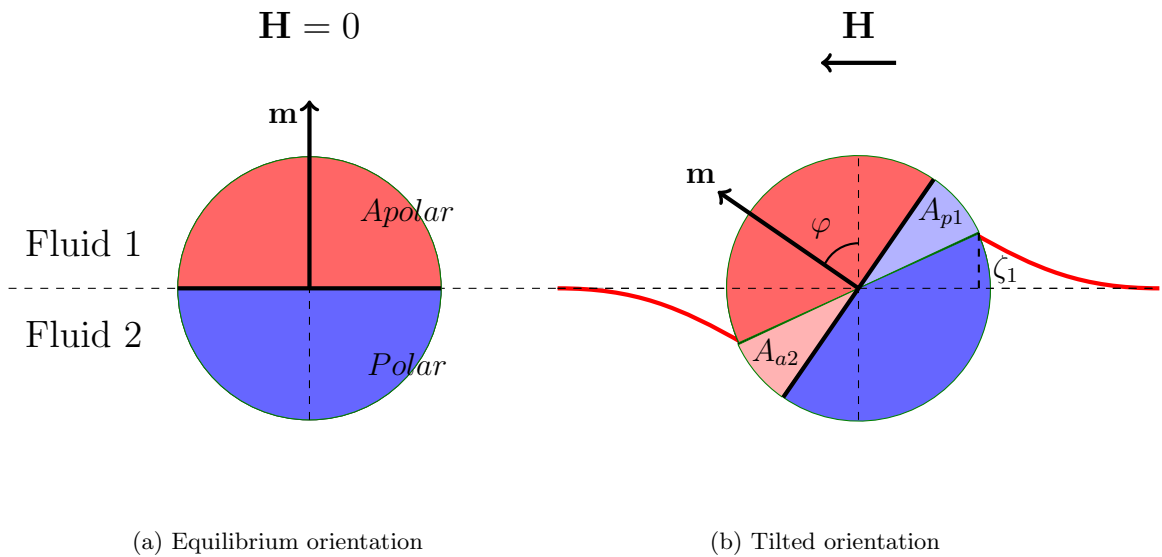
\subsection{The colloidal particle}
\label{ssec:nanoparticles}
The trajectory of the colloidal Janus particle is updated using a leap-frog integrator.
The particle is discretized on the fluid lattice and coupled to the fluid species by means of 
a modified bounce-back boundary condition as pioneered by Ladd and Aidun~\cite{ladd-verberg2001, AIDUN1998}.

The outer shell of the particle is filled with a ``virtual'' fluid with the density
\begin{eqnarray}
 \rho_{\mathrm{virt}}^1(\vec{x},t)  &=& \overline{\rho}^1(\vec{x},t) + |\Delta\rho| \mbox{, } \\
 \rho_{\mathrm{virt}}^2(\vec{x},t) & =& \overline{\rho}^2(\vec{x},t) - |\Delta\rho|  \mbox{, }
\end{eqnarray}
where $\overline{\rho}^1(\vec{x},t)$ and $\overline{\rho}^2(\vec{x},t)$ are the average of the density of 
neighbouring fluid nodes for component $1$ and $2$, respectively. 
The parameter $\Delta\rho$ is called the ``particle colour'' and dictates the contact angle of the particle.
A particle colour $\Delta\rho = 0$ corresponds to a contact angle of $\theta_p = 90^{\circ}$, i.e. a neutrally wetting particle. 
In order to simulate a Janus particle, we set different particle colours in well defined surface areas corresponding to the different hemispheres of the particle.

We choose a system size $S = 128 \times 64 \times 128$ to eliminate finite size
effects. We fill one half of the system with fluid $1$ and the other half with
fluid $2$ of equal density ($\rho_1=\rho_2 = 0.7$) such that a fluid-fluid
interface forms at $y=32$. The interaction strength in~\eqnref{sc} is chosen to
be $g_{12} = 0.1$ and the particle with radius $R =10$ is placed at the
interface.  We impose walls with mid-grid bounce back boundary conditions at
the top and bottom of the system parallel to the interface, while all other
boundaries are periodic.

\section{Theoretical Results}

\label{sec:theory}
We consider a spherical Janus particle composed of apolar and polar hemispheres
adsorbed at a fluid-fluid interface, as illustrated in \figref{equi}.  The two
hemispheres have opposite wettability, represented by the three-phase contact
angles $\theta_A = 90^{\circ} +\beta$  and $\theta_P = 90^{\circ} -\beta$,
respectively, where $\beta$ represents the amphiphilicity of the particle. A
larger $\beta$ value corresponds to a greater degree of particle
amphiphilicity.  In its equilibrium state, the Janus particle takes an upright
orientation with respect to the interface with its two hemispheres totally
immersed in their favourable phases, as shown in \figref{equi}.

The free energy of the particle in its equilibrium configuration is
\begin{equation}
 E_{\text{int}} = \gamma_{12}A_{12}^{ \text{int}} + \gamma_{a1}A_{a1}^{\text{int}} +\gamma_{p2}A_{p2}^{\text{int}}
  \mbox{,}
  \end{equation}
where $\gamma_{ij}$ are the interface tensions between phases $i$ and $j$ and $A_{ij}$ are the
contact surface areas between phases $i$ and $j$, where $i,j$ $=$ $\{1$: fluid, $2$: fluid, $a$: apolar, $p$: polar$\}$.
For a symmetric amphiphilic spherical particle, the apolar and polar surface
areas are equal $A_{a1}=A_{p2} = 2\pi R^2$.

After switching on the horizontal magnetic field, $\mathbf{H}$, the particle experiences a torque $\boldsymbol{\tau}=\mathbf{m} \times \mathbf{H}$ that attempts to align the particle dipole axis with the field. However, surface tension resists the rotation causing the particle to tilt with respect to the interface for a given dipole-field strength $B=|\mathbf{m}| |\mathbf{H}|$, as illustrated in \figref{def}. The tilt angle $\varphi$ is defined as the angle between particle dipole-moment and the undeformed interface normal (i.e. the $y$-axis).
The interface deforms around the particle so that each fluid contacts a larger area of its favourable particle surface (\figref{def}).
This interface deformation increases the fluid-fluid interface area and decreases the surface of each hemisphere contacting its unfavourable fluid.
The free energy of the system is reduced in total 
due to the dominant contribution of particle-fluid interface energies~\cite{Rezvantalab2013}.
\begin{figure*}[!t]
\begin{subfigure}{.405\textwidth}
\includegraphics[width= 0.95\textwidth]{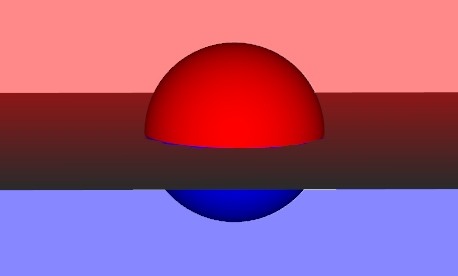}
\caption{}
\label{fig:janus-geo-init}
\end{subfigure}
\begin{subfigure}{.4\textwidth}
\includegraphics[width= 1.0\textwidth]{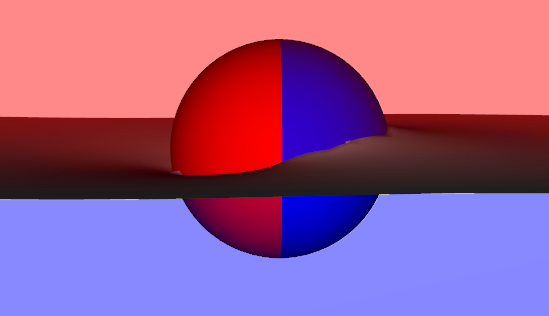}
\caption{}
\label{fig:janus-geo-tilt}
\end{subfigure}
\caption{Snapshots of a Janus particle at a fluid-fluid interface 
at $(a)$ initial equilibrium state ($\varphi = 0^{\circ}$)  and  $(b)$ tilted orientation state ($\varphi = 90^{\circ}$)
as obtained from our simulations.
The three-phase contact line undulates around the tilted particle so that the interface is deformed.}
\captionsetup[figure]{slc=off}
\label{fig:janus-geo}
\end{figure*}
The free energy of a Janus particle in a tilted orientation can be written as
\begin{eqnarray}
 E_{\text{tilt}} = &\gamma_{12}A_{12}^{\text{tilt}} + \gamma_{a1}A_{a1}^{\text{tilt}} + \gamma_{p2}A_{p2}^{\text{tilt}} \nonumber \\ 
   +&\gamma_{a2}A_{a2} + \gamma_{p1}A_{p1} + B \sin\varphi
  \mbox{.}
\end{eqnarray}
The free energy difference between the tilted orientation state and the initial state is
\begin{eqnarray}
 \Delta E & = & E_{\text{tilt}} - E_{\text{int}} \nonumber \\ 
          & = & \gamma_{12} \left( A_{12}^{\text{tilt}} -A_{12}^{ \text{int}} \right) +  \gamma_{a1} \left( A_{a1}^{\text{tilt}} - A_{a1}^{\text{int}}  \right) \nonumber \\ 
          & & +  \gamma_{p2}\left(A_{p2}^{\text{tilt}} - A_{p2}^{\text{int}} \right) + \gamma_{a2}A_{a2} +\gamma_{p1}A_{p1} \nonumber \\
          & & + B \sin\varphi
          \mbox{.}
\end{eqnarray}
Since $ A_{a1}^{\text{int}} = A_{a1}^{\text{tilt}} + A_{a2}$ and $ A_{p2}^{\text{int}} = A_{p2}^{\text{tilt}} +A_{p1} $, we obtain 
\begin{eqnarray}
 \Delta E  &=& \gamma_{12}\Delta A_{12}  +(\gamma_{a2}-\gamma_{a1}) A_{a2} \nonumber \\ 
         & & + (\gamma_{p1}-\gamma_{p2})A_{p1} + B \sin\varphi
   \mbox{,}
 \label{eq:free}
\end{eqnarray}
where $\Delta A_{12} = A_{12}^{\text{tilt}} -A_{12}^{\text{int}}$ is the increased fluid-fluid interface area.
The particle obeys Young's boundary conditions~\cite{Ondurcuhu1990}
\begin{equation}
 \cos \theta_A = \frac{\gamma_{a1}-\gamma_{a2}} { \gamma_{12}} \mbox{, } \qquad \cos \theta_P = \frac{\gamma_{p1}-\gamma_{p2}} {\gamma_{12}}
 \mbox{.}
\end{equation}
For two hemispheres with opposite wettabilities, we obtain $\cos \theta_A = -\cos \theta_P = -\sin \beta$.
\eqnref{free} is further simplified to
\begin{equation}
  \Delta E = \gamma_{12} \Delta A_{12}  +  \gamma_{12} (A_{a2} + A_{p1}) \sin \beta + B \sin\varphi
    \mbox{.}
 \label{eq:free_easy}
\end{equation}
Under the assumption of a flat interface~\cite{Park2012a}, $\Delta A_{12}=0$, $A_{a2} =A_{p1}= \frac{\varphi}{2\pi} 4\pi R^2= 2\varphi R^2 $. 
Therefore, \eqnref{free} finally reduces to
\begin{equation}
\Delta E = 4 \varphi R^2 \gamma_{12} \sin \beta + B \sin\varphi
\mbox{.}
 \label{eq:eflat}
\end{equation}

There is no exact analytical expression for the free energy of a tilted Janus particle at an interface that includes interface deformations, due to the difficulty in modelling the shape of the interface and position of the contact line. 
However, in the limit of small interface deformations~\cite{Stamou2000}, we derive such an analytical expression for the free energy.

We consider micron-sized particles with a radius much smaller than the capillary length such that we can neglect the effect of gravity.
We assume the pressure drop across the interface to be zero, leading to vanishing mean curvature according to the Young-Laplace equation.
The mean curvature can be approximately written in cylindrical coordinates as~\cite{Stamou2000}
\begin{equation}
\label{eq:multipole_solution}
 \Delta h (r,\vartheta) = \left(\frac{1}{r}\frac{\partial}{\partial r} r \frac{\partial}{\partial r}+\frac{1}{r^2}\frac{\partial}{\partial \vartheta^2}\right) h(r,\vartheta) = 0
 \mbox{,}
\end{equation}
where $h$ is the height of the interface. The radial distance $r$ and the polar angle $\vartheta$ are defined with respect to a particle centred reference frame. Using a multipole analysis~\cite{Stamou2000}, the solution of Eq.~\eqref{eq:multipole_solution} yields 
\revisedtext{
\begin{align}
 h (r,\vartheta) = & \sum_{m>0} R_m(r)\Phi_m (\vartheta) \\
  = & \sum^\infty_{m>0} \zeta_m \cos\left(m(\vartheta-\vartheta_{m,0})\right) \left( \frac{r_c}{r} \right)^m
 \mbox{,}
\end{align}
where $\zeta_m$ is the maximal height of the contact line,} $\vartheta_{m,0}$ is the phase angle and $r_c$ is the radius 
where the particle and fluid interface intersect. $r_c$ is approximately the particle radius 
in the limit of small interface deformations.
\revisedtext{The monopolar term $m=0$ is ommitted because we focus on micron-sized particles where gravitational effects can be neglected.} 
The dipolar term $m=1$, results from an external torque on the particle, causing symmetric interface rise and depression around it.
$m=2$ denotes the quadrupole term, which dominates in the absence of any external forces or torques on the particle.
\revisedtext{ 
We consider only the leading order $m=1$ dipole term}
\begin{equation}
 h (r,\vartheta) = \zeta_1 \cos\left(\vartheta-\vartheta_{1,0}\right) \frac{r_c}{r}
 \mbox{,}
\end{equation}
where $\zeta_1$ is the maximal height of the contact line.
We calculate the increased fluid-fluid interface area $\Delta A_{12}$ by considering an infinitesimal element $dA^*=dx^*\times dy^*$ of the deformed interface.
In a local coordinate system rotated such that the slope is maximized along the $y$ coordinate, we have~\cite{Stamou2000}
\begin{eqnarray}
 dx^* & = & dx \mbox{,} \\ 
 dy^* & = & \sqrt{dy^2+dh^2} \approx dy\left(1+\frac{1}{2}(\nabla h)^2\right)   
 \mbox{,}
\end{eqnarray}
resulting in
\begin{eqnarray}
  d (\Delta A_{12}) &=& dx^*\times dy^* - dx\times dy \nonumber \\
    & & =  dx\,dy\frac{1}{2}\left(\nabla h\right)^2
\end{eqnarray}
or
\begin{equation}
 \Delta A_{12} = \frac{1}{2} \int_{r=r_c}^{\infty} \int_{\vartheta=0}^{2\pi} {(\nabla h)^2 r d\vartheta dr}
 \mbox{.}
\end{equation}
In the dipole approximation, one has 
\revisedtext{
\begin{eqnarray}
(\nabla h) \cdot (\nabla h) &=& \left(\frac{\partial}{\partial r} h\right)^2  + \frac{1}{r^2} \left( \frac{\partial}{\partial \vartheta} h\right) ^2  \nonumber \\
& = & \zeta_1^2 r_c^2r^{-4} 
\end{eqnarray}
}
and we obtain 
\begin{equation}
 \Delta A_{12} =  \frac{\pi}{2} \zeta_1^2 
 \mbox{.}
\end{equation}
The areas $A_{a2}$ and $A_{p1}$ can then be written as  
\begin{eqnarray}
 A_{a2}= A_{p1} & =&  2 R^2\varphi - \int_{0}^{\pi} {h(r=r_c) r_c d\vartheta} \nonumber \\
 &=& 2 R^2\varphi - 2 r_c \zeta_1 
 \mbox{.}
\end{eqnarray}
We assume $r_c = R$ and the free energy taking into account small interface deformations can then be written as
\begin{equation}
 \Delta E = 
  \frac{\pi}{2} \gamma_{12} \zeta_1^2 + 4 ( R^2\varphi- R \zeta_1) \gamma_{12} \sin \beta + B \sin\varphi
  \mbox{.}
   \label{eq:edef}
\end{equation}
We discuss the predictions of our models and compare them with our simulation results in the following section. 

\section{Simulation results and comparison to theory} 

\begin{figure}[h]
\includegraphics[width= 0.5\textwidth]{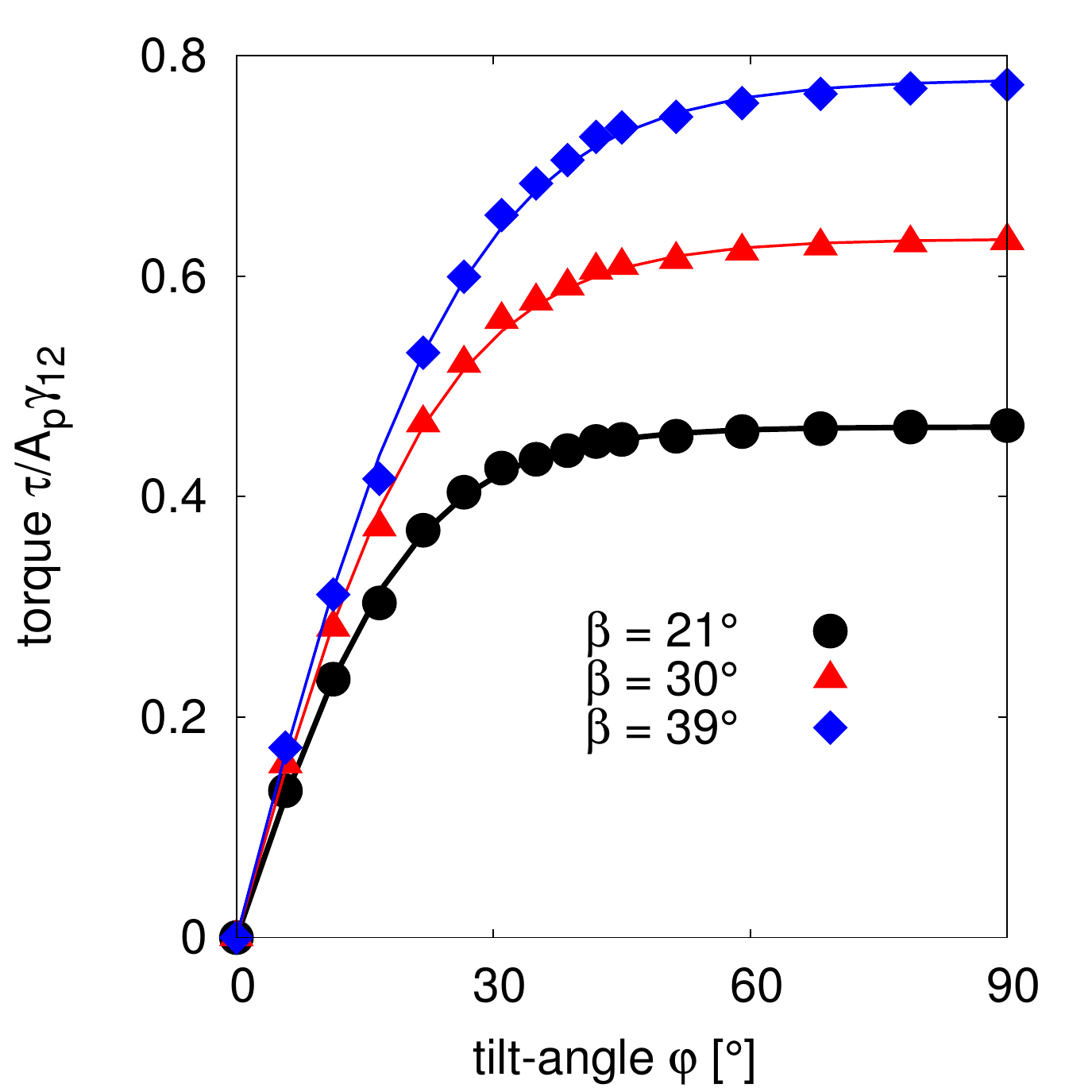}
\caption{Reduced torque $\tau/A_p\gamma_{12}$ as a function of tilt angle $\varphi$ for  $\beta = 21^{\circ}$ (circles) , $\beta = 30^{\circ}$ (triangles) and $ \beta = 39^{\circ}$ (diamonds), where $A_p = \pi R^2$. 
The symbols are simulation data and the solid lines represent hyperbolic tangent functions to fit the data. 
The fitted functions are integrated in order to obtain the free energy. For all amphiphilicities, the torque is linear for small
rotations around the equilibrium position, before reaching a constant value at very large tilt angles $\varphi \to 90^{\circ}$.
}
\label{fig:angle_tor}
\end{figure}
\label{sec:simu}

In addition to the theoretical models we developed in Section~\ref{sec:theory}, we numerically calculate the free energy of a single Janus particle adsorbed at a fluid-fluid interface using lattice Boltzmann simulations.
Our lattice Boltzmann simulations are capable of capturing interface deformations fully without making any assumptions about the magnitude of the deformations or stipulating any particle-fluid boundary conditions. 

\figref{janus-geo-init} shows the initial equilibrium configuration of our Janus particle simulation, where each hemisphere totally immerses in its corresponding favourable liquid so that the interface remains flat.
\figref{janus-geo-tilt} shows how the three-phase contact line and interface deform around the Janus particle as it tilts with respect to the interface.

In order to obtain the free energy of the Janus particle as a function of tilt angle from our simulations,
the total surface area of the deformed fluid-fluid interface and the corresponding particle surfaces have to be measured~\cite{Rezvantalab2013,gompper2014,Gary2014a}. In this paper, we employ a simple method, which utilises the ability to easily measure the force applied to the particle by the fluids using the lattice Boltzmann method. 

We first determine the contribution to the free energy neglecting the dipole-field contributions by integrating \revisedtext{the torque} on the particle as the particle rotates quasi-statically,

\begin{equation}
\label{eq:tor_eq}
 \Delta E = \int_{0}^{\varphi_{\text{tilt}}} \tau_{\varphi} d\varphi
 \mbox{,}
\end{equation}
where $\varphi_{\text{tilt}}$ is the tilt angle of interest. 
To do this, we rotate the particle on the interface until it reaches the desired tilt angle and then fix the position of the particle and let the system equilibrate. The remaining torque on the particle is the resistive torque applied to the particle from the fluid-fluid interface. 

\figref{angle_tor} shows the evolution of this torque $\tau_{\varphi}$ versus the tilt angle for different amphiphilicities $\beta = 21^{\circ}$ (circles), $\beta = 30^{\circ}$ (triangles), and $ \beta = 39^{\circ}$ (diamonds), corresponding to particle colours $\Delta\rho = 0.10$, $0.15$, $0.20$, respectively.
For all amphiphilicities, the torque increases linearly as the particle rotates for small tilt angles, $\varphi < 30^{\circ}$. As the tilt angle increases further $\varphi \to 90^{\circ}$ the torque tends to a nearly constant value.
We fit the torque $\tau_{\varphi}$ with a hyperbolic tangent function, and integrate the fitted function numerically to obtain the free energy.

\begin{figure}[!t]
\centering
\includegraphics[width= 0.5\textwidth]{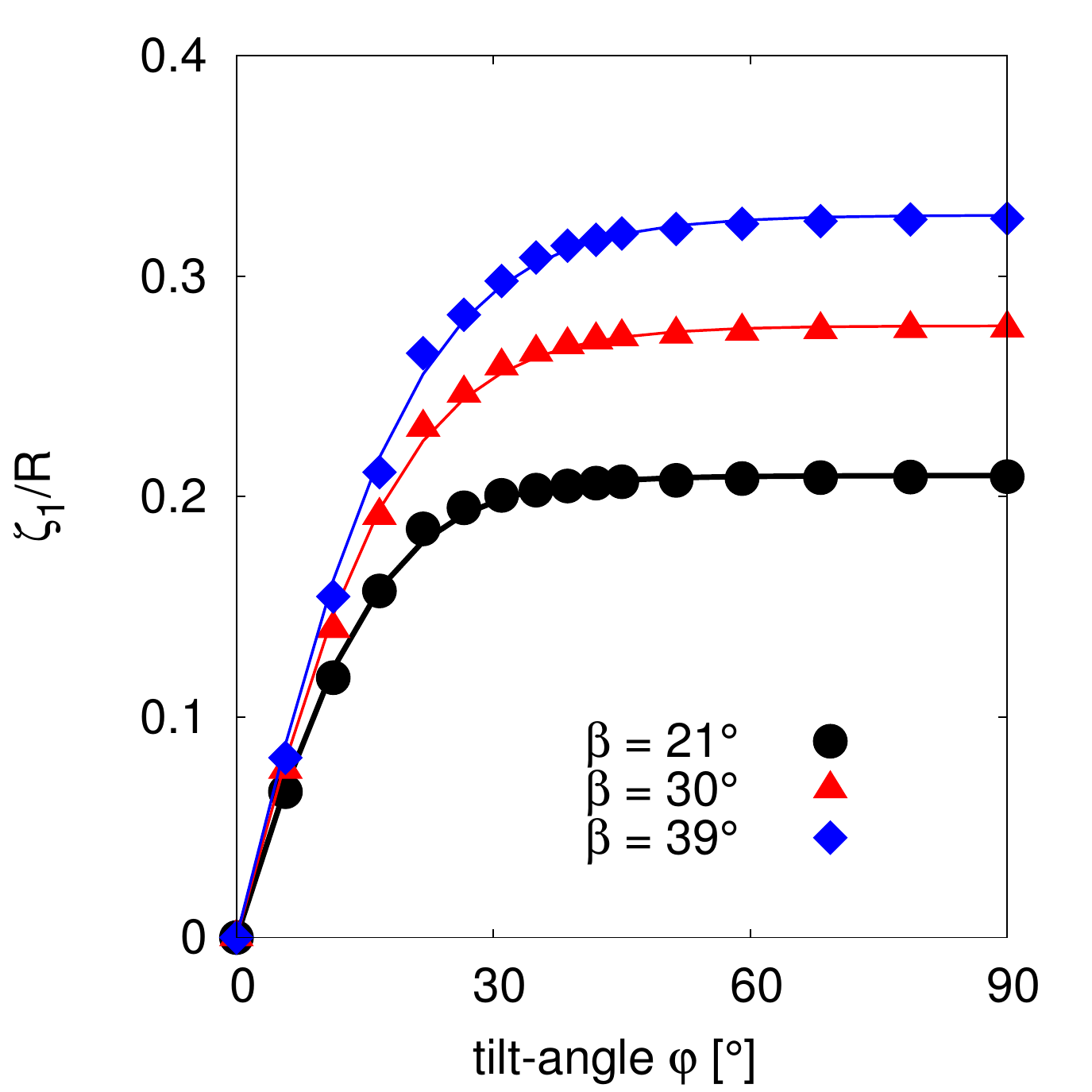}
\caption{Reduced maximal height of the deformed interface $\zeta_1/R$ as a function of tilt angle $\varphi$ for  $\beta = 21^{\circ}$(circles) , $\beta = 30^{\circ}$ (triangles) and $ \beta = 39^{\circ}$ (diamonds).
The data (symbols) are fitted with a hyperbolic tangent function (solid lines). The maximal interface height increases linearly with tilt angle for small angles, and becomes constant for large angles.}
\label{fig:height_angle}
\end{figure}

In order to calculate the free energy of our small interface deformation based model (\eqnref{edef}),
we measure the corresponding maximal height of the contact line $\zeta_1$ as a function of tilt angle, as shown in \figref{height_angle}. 
Similarly to \figref{angle_tor}, the height of the contact line increases linearly for small tilt angles, and then reaches a plateau for large tilt angles. 
The height plateau demonstrates that the deformed interface area remains constant at large tilt angles.
This indicates that at large tilt angles, only the particle-fluid surface energy, which increases linearly with increasing tilt angle, contributes to the change in free energy, and therefore the torque $\tau \propto \,d (\Delta E)/ \,d \varphi$ is constant at large tilt angles, in agreement with \figref{angle_tor}.

\begin{figure}[bhtp]
\includegraphics[width= 0.5\textwidth]{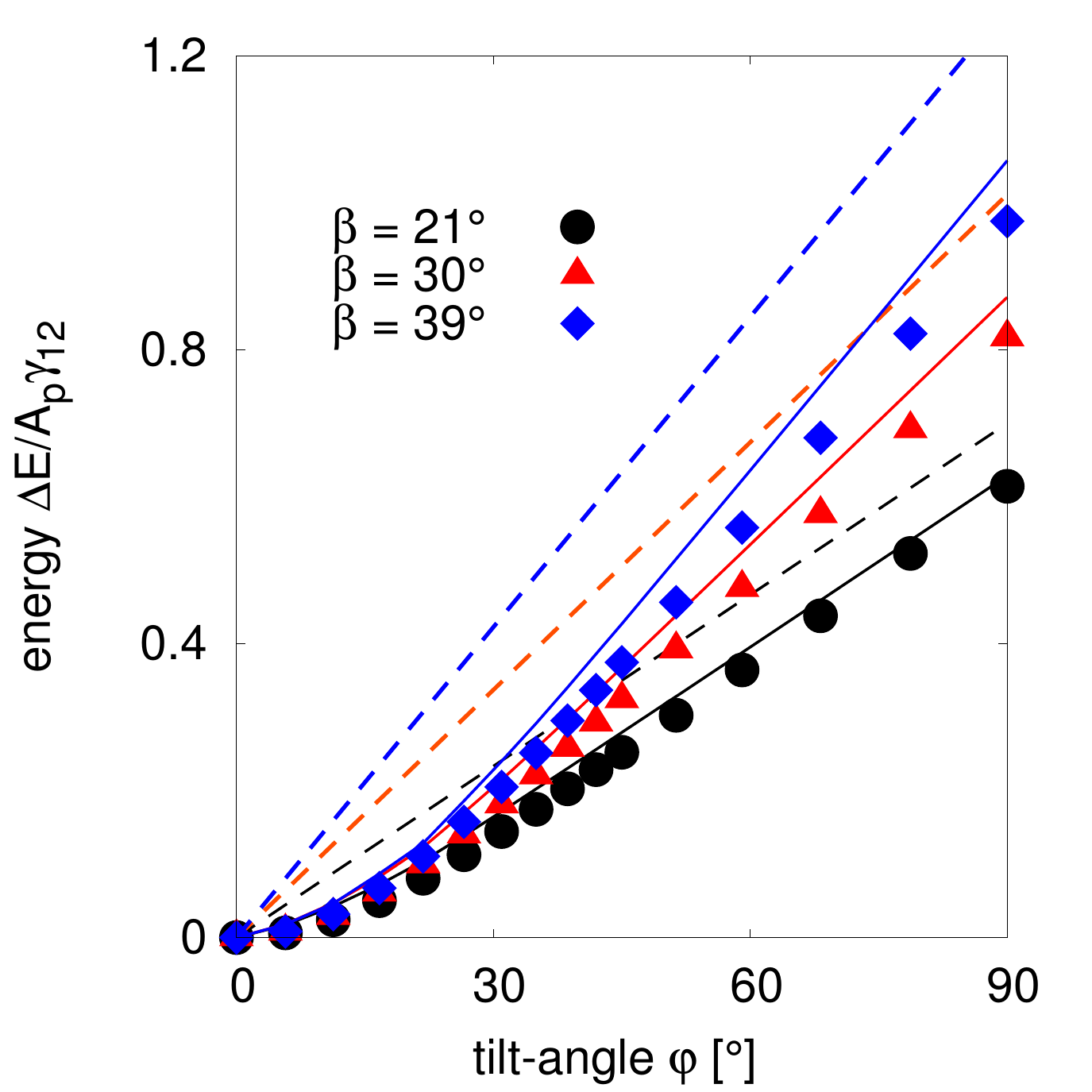}
\caption{Free energy as a function of tilt angle $\varphi$ for  $\beta = 21^{\circ}$ (circles), $\beta = 30^{\circ}$ (triangles) and $ \beta = 39^{\circ}$ (diamonds), 
which are calculated using the analytical model that excludes interface deformations (\eqnref{eflat}, dashed lines),
our analytical model which assumes small interface deformations (\eqnref{edef}, solid lines) 
and numerically integrated results (symbols) without considering a magnetic dipole-field contribution ($B=0$).
The undeformed interface model \eqnref{eflat} shows non-negligible deviation from the simulation data.
For small $\beta$ and small $\varphi$, the small deformation model \eqnref{edef} is in good agreement with the simulation results.
}
\label{fig:angle_energy}
\end{figure}

In \figref{angle_energy} we compare the free energy models that we developed in Section~\ref{sec:theory} assuming no interface deformations (\eqnref{eflat}, dashed lines) 
and small interface deformations (\eqnref{edef}, solid lines) with our lattice Boltzmann simulation data (symbols), which incorporates interface deformation fully, by measuring the free energies as a function of particle tilt angle as described above.

Our undeformed interface model (\eqnref{eflat}) predicts that the energy varies linearly with the particle tilt angle, $\varphi$,
for all particle amphiphilicities, $\beta = 21^{\circ}$, $30^{\circ}$, and $39^{\circ}$. The model also predicts that the free energy 
increases as the amphiphilicity increases from $\beta = 21^{\circ}$ to $\beta = 39^{\circ}$ for any given tilt angle.
\begin{figure}[!t]
\centering
\includegraphics[width= 0.5\textwidth]{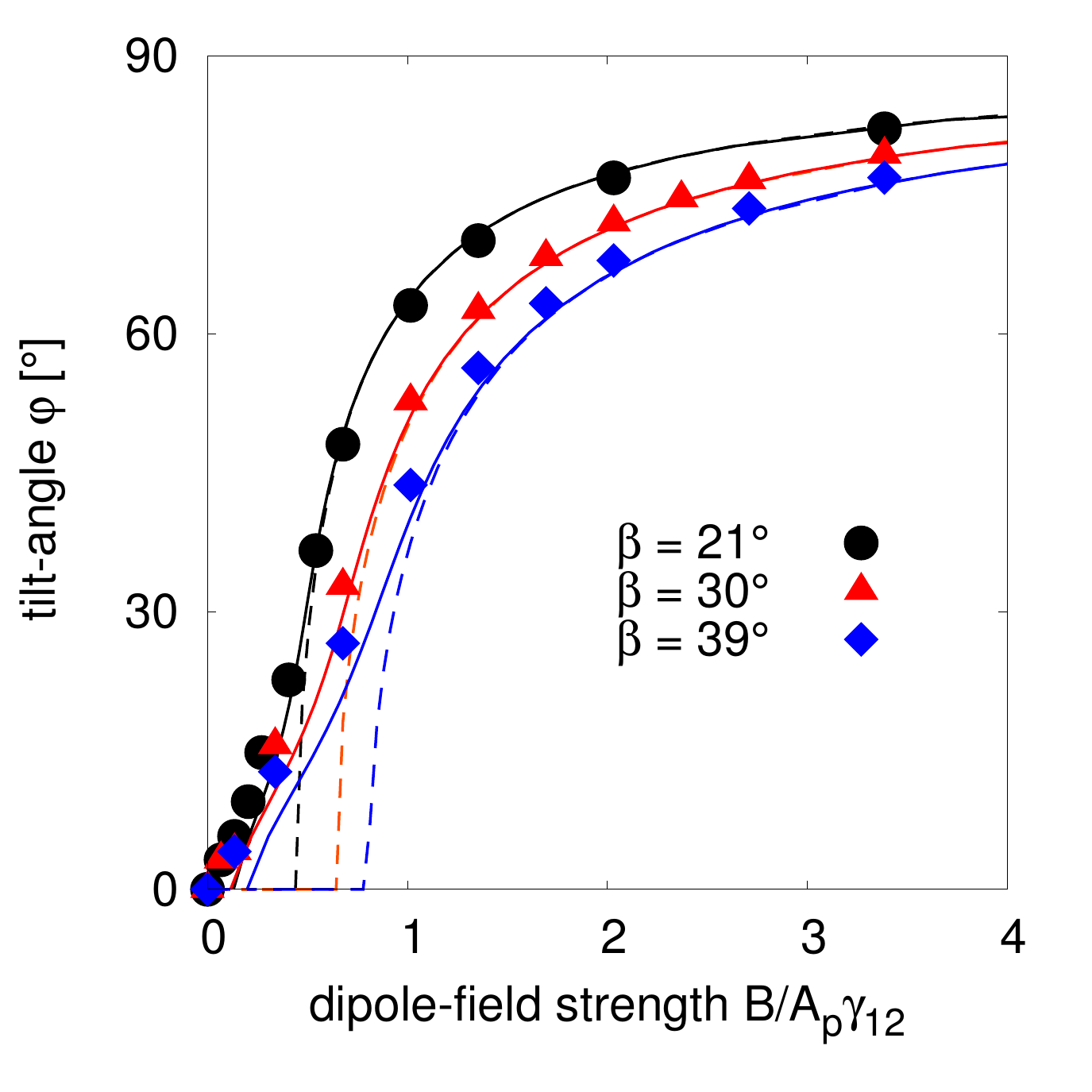}
\caption{Tilt angle $\varphi$ as a function of dipole-field strength for different amphiphilicities $\beta = 21^{\circ}$(circles), $\beta = 30^{\circ}$ (triangles) and $ \beta = 39^{\circ}$ (diamonds). 
We compare the simulation data (symbols) with the tilted angles predicted using the analytical model which excludes interface deformations (\eqnref{eflat}, dashed lines),
and our analytical model which assumes small interface deformations (\eqnref{edef}, solid lines).
For small dipole-field strengths, the tilt angle predicted by the undeformed interface model shows large deviations from the simulation data,
whereas the tilt angle obtained from the small deformation model agrees well with the simulations.
}
\label{fig:angle_dip}
\end{figure}
Our analytical model assuming small interface deformations (\eqnref{edef}) shows some interesting qualitative behaviour. 
Firstly, for small tilt angles, $\varphi < 30^{\circ}$, the energy varies quadratically with the tilt angle for all amphiphilicities.
This is because for small tilt angles, the maximal interface height $\zeta_1$ varies linearly with the tilt angle (\figref{height_angle}), $\zeta_1 \approx R\varphi$, and the free energy in \eqnref{edef} becomes approximately $\Delta E = \frac{\pi}{2}\gamma_{12}R^2\varphi^2$.

Secondly, for larger tilt angles, $\varphi > 45^{\circ}$, the energy varies linearly with the tilt angle, due to the fact that for large angles, $\zeta_1$ is constant (\figref{height_angle}). Therefore,
\eqnref{edef} becomes $\Delta E = 4R^2\gamma_{12} \sin\beta \varphi + C$, where $C$ is a constant, which explains the linear behaviour of the free energy for large tilt angles.

The above analysis also explains why, for tilt angles $\varphi < 30^{\circ}$, the energy only weakly depends on the amphiphilicity $\beta$ of the particles:
the amphiphilicity term $\sin \beta$ is only significant for large tilt angles.
  
When comparing these two models with our simulation data (symbols), we find that the small deformation model 
captures the qualitative features of the data extremely well. In addition, the model quantitatively agrees with the numerical 
results for small tilt angles $\varphi < 30^{\circ}$ for all amphiphilicities. As the tilt angle of the particles increases the 
quantitative deviation between the model and the data becomes more significant. However, for small amphiphilicities $\beta = 21^{\circ}$ the model and numerical data are in good agreement.

In contrast, our theoretical model that takes into account only the free energy differences between the particle as a function of its orientation and therefore neglects interface deformations (\eqnref{eflat}) performs much worse. The model captures the qualitative linear behaviour of 
the numerical data only for large tilt angles $\varphi > 50^{\circ}$, but with large quantitative differences that increase as the particle amphiphilicity increases.
The results in \figref{angle_energy} clearly show that interface deformations strongly affect the behaviour of a tilted Janus particle adsorbed at a fluid-fluid interface, and we note that our 
analytical model that includes small interface deformations is clearly able to capture this qualitative behaviour. 

To verify the predictions of the analytical models including dipole-field contributions, we switch on the magnetic field and numerically measure the time-averaged tilt angle of the particle after the system has equilibrated, as per \eqnref{tor_eq}.
We obtain the tilt angle as a function of the dipole-field strength by minimizing the free energies in \eqnref{eflat} and \eqnref{edef} with respect to the tilt angle.

\begin{figure}[!t]
\includegraphics[width= 0.45\textwidth]{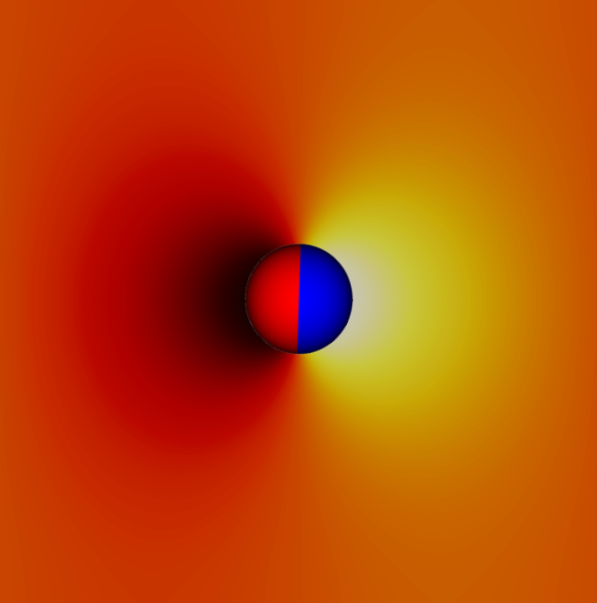}
\caption{Plot of the relative height of the interface. Influenced by a magnetic torque, the Janus particle reorients, leading to dipolar interface deformations: the interface
is depressed on one side of particle (black) and raised on the other side (yellow). 
The interface is flat in the orange/red regions. These dipolar interface deformations are dynamically tunable, 
providing a route to generate tunable capillary-driven assembly at fluid interfaces.
}
\label{fig:depth}
\end{figure}
In \figref{angle_dip}, we compare the predicted tilt angles from 
the free energy models assuming no interface deformation (\eqnref{eflat}, dashed lines) 
and small interface deformation (\eqnref{edef}, solid lines) with our numerical simulation data (symbols), which incorporates interface deformation 
fully, by measuring the tilt angle as a function of dipole-field strength, for $\beta = 21^{\circ}$, $30^{\circ}$, and $\beta = 39^{\circ}$.

For large dipole-field strengths $B > 1.5 A_p \gamma_{12}$, both analytical models perform well by predicting the numerically measured tilt angles for all particle amphiphilicities $\beta$.
This is due to the fact that large dipole-field strengths cause large tilt angles at which the deformed interface area stops increasing, as shown in \figref{height_angle}. In this regime, the equilibrium orientation only depends on the particle properties (radius $R$ and amphiphilicity $\beta$), interface tension $\gamma_{12}$, and dipole-field strength $B$, which are incorporated into both analytical models.

For small dipole-field strengths, the tilt angle predicted by the undeformed interface model \eqnref{eflat} shows large deviations from the simulation data for all amphiphilicities.
The deviations increase with increasing particle amphiphilicity.
In addition, the undeformed interface model predicts a large critical dipole-field strength at which the particle begins to rotate, which we do not observe in the simulations.
This critical dipole-field strength increases with increasing particle amphiphilicity. From \eqnref{eflat}, we can obtain the torque, $\tau \propto \frac{d (\Delta E)}{d\varphi} = 4R^2\gamma_{12} \sin\beta + B \cos \varphi$ where the first term is independent of the tilt angle $\varphi$. Therefore, in the zero dipole-field state $\varphi = 0 ^{\circ}$, a critical dipole-field strength $B_c = 4R^2\gamma_{12} \sin\beta$ is needed to overcome this constant resistive component of the torque.

Our small deformation model (\eqnref{edef}) shows significant improvement at
predicting the tilt angle compared with the undeformed interface model. The
predicted tilt angle from this model shows good agreement with simulation data
for weak dipole-field strengths, especially for amphiphilicity
$\beta=21^{\circ}$.  In particular the model predicts a much smaller, though
still finite, critical dipole-field strength, agreeing better with our
numerical simulations. We conclude that interface deformations strongly affect
the orientation of a tilted Janus particle adsorbed at a fluid-fluid interface. 

Finally, ~\figref{depth} shows that the Janus particle deforms the interface in
a dipolar fashion: a symmetric depression and rise on opposite sides of the
particle, as also shown in~\figref{janus-geo-tilt}. Since the strength of the
capillary interactions between two particles interacting as polar capillary
dipoles depends on the maximal interface deformation height $\zeta_1$, we can
tune the strength of capillary interactions by varying the dipole-field
strength. These unique dipolar capillary interactions may be used to assemble
particles into novel materials at a fluid-fluid interface in a tunable
way~\cite{Gary2014b}.

\section{Conclusion}
\label{sec:final}
We studied the behaviour of a magnetic spherical amphiphilic Janus particle
adsorbed at a fluid-fluid interface under the influence of an external magnetic
field directed parallel to the interface.

The interaction of the particle with the external field results in a torque
that tilts the particle, introducing interface deformation. We derived
analytical models assuming no interface deformations and small interface
deformations that enable the calculation of the free energy, $\Delta E$, and
hence the equilibrium orientation $\varphi$, of the particle in terms of the
particle size $R$, particle amphiphilicity $\beta$, fluid-fluid interface
tension $\gamma_{12}$, maximal interface height $\zeta_1$, and magnetic
dipole-field strength $B$. We used lattice Boltzmann simulations that
incorporate interface deformations fully to test the results of our analytical
models. 

In the absence of a magnetic field, our simulations showed that the maximal
interface deformation height $\zeta_1$ increases linearly $\zeta_1 \approx
R\psi$ for small tilt angles and then reaches a plateau for large tilt angles
$\zeta_1 \approx \mathrm{const}$. They also showed that the free energy of the
particle increases quadratically for small tilt angles and linearly for large
tilt angles. We explain this behaviour in terms of our theoretical model
assuming small interface deformations that correctly predicts this behaviour,
and find that the free energy varies as $\Delta E =
\frac{\pi}{2}\gamma_{12}R^2\varphi^2$ for small tilt angles and as $\Delta E =
4R^2\gamma_{12} \sin\beta \varphi + C$ for large tilt angles. 

With the magnetic field switched on, we calculated the dependence of the tilt
angle $\psi$, on the dipole-field strength $B$, by minimising the calculated
free energies with respect to the tilt angle $\psi$, for a given dipole-field
strength $B$. 

Our undeformed interface model based on free energy differences agrees
qualitatively with our simulation results only for large particle tilt angles,
but deviates significantly quantitatively. In particular, this model predicts a
critical dipole-field strength $B_c = 4R^2\gamma_{12} \sin\beta$ in order to
rotate the particle, which we do not observe in our simulations. Our results
contrast with other free energy difference based models that assume a flat
interface, which were able to predict the equilibrium orientation of
nanoparticles reasonably accurately~\cite{Park2012a,Park2012c}.

Our analytical model that includes small interface deformations captures the
qualitative behaviour of the particle well for all tilt angles, and also
performs well quantitatively, in particular for small particle amphiphilicities
$\beta$. Previous studies by Davies et al.~\cite{Gary2014a} have shown that
interface deformations can significantly alter the quantitative behaviour of
ellipsoidal particles at interfaces, however, our results demonstrate that
interface deformations can both \emph{quantitatively and qualitatively} affect
the behaviour of Janus particles at fluid-fluid interfaces and therefore cannot
be neglected.

Finally, we show that the interface deformations around a spherical magnetic
titled Janus particle influenced by an external field directed parallel to the
interface are dipolar in nature.
Further, we demonstrate that the magnitude of these deformations can be
dynamically tuned, and therefore the capillary interactions between monolayers
of such particles are tunable, and we suggest that this may enable the
production of novel, reconfigurable materials.

\begin{acknowledgments}
Financial support is acknowledged from NWO/STW (Vidi grant 10787 of J.~Harting
and STW project 13291). We thank the J\"ulich Supercomputing Centre and the
High Performance Computing Center Stuttgart for the technical support and the
CPU time which was allocated within grants of the Gauss Center for
Supercomputing and the ``Partnership for Advanced Computing'' (PRACE). GBD
acknowledges Fujitsu and EPSRC for funding. 
\end{acknowledgments}

\end{document}